\documentclass[epj]{svjour}
 \usepackage[english]{babel}
 \usepackage{amsmath}
 \usepackage{amssymb}
   \usepackage{graphicx,color}

\def\E{\mathcal E}
\def\R{\mathcal R}
\let\e=\varepsilon
\let\d=\delta

\newcommand{\be}{\begin{equation}}
\newcommand{\ee}{\end{equation}}
\newcommand{\ba}{\begin{eqnarray}}
\newcommand{\ea}{\end{eqnarray}}
\newcommand{\nn}{\nonumber \\}

\begin{document}

\title{The symmetric 1:2 resonance}
\author{A.~Marchesiello\inst{1}
\and
G.~Pucacco
\inst{2}\thanks{pucacco@roma2.infn.it.}}
\institute{Dipartimento di Scienze di Base e Applicate per l'Ingegneria -- Universit\`a di Roma ``la Sapienza",
Via Antonio Scarpa, 16 - 00161 Roma
\and Dipartimento di Fisica -- Universit\`a di Roma ``Tor Vergata" and INFN -- Sezione di Roma Tor Vergata, Via della Ricerca Scientifica, 1 - 00133 Roma}

\date{Received: date / Revised version: date\\
Published online: date}

%

\abstract{This paper illustrates the application of Lie transform normal-form theory to the construction of the 1:2 resonant normal form corresponding to a wide class of natural Hamiltonian systems. We show how to compute the bifurcations of the main periodic orbits in a potential with double reflection symmetries. The stability analysis of the normal modes and of the periodic orbits in general position allows us to get overall informations on the phase-space structure of systems in which this resonance is dominating. As an example we  apply these results to a class of models useful as galactic potentials.
}

\maketitle

\section{Introduction}

The 1:2 resonance or `Fermi resonance' plays a prominent role in nonlinear Hamiltonian dynamics. In galactic dynamics it appears in several fashions \cite{fv}: to mention a few, in axisymmetric prolate systems it determines the bifurcation of the {\it inner thin tube} orbits \cite{eva}; in triaxial systems with ellipsoidal non-singular equipotentials it gives the bifurcation of {\it banana} and {\it anti-banana} orbits in the symmetry planes \cite{Bi,mes}. Its interest is clearly not limited to this field and its investigation in theoretical and applied nonlinear dynamics has been very active \cite{dui,CDHS}: an example is the so-called {\it spring-pendulum} \cite{Br2:1} and an application in satellite attitude dynamics is the {\it tethered system} \cite{SC}; in chemistry it is quite relevant in molecular vibrations \cite{joy1,joy2} and in quantum physics for the semi-classical approximation of atomic nuclei \cite{ab}.

In \cite{MP} we have investigated the relevance of the 1:1 resonance in galactic dynamics in the cases of one and two reflection symmetries. Here we want to show how Hamiltonian  normal forms can be used to get qualitative and {\it quantitative} information on the bifurcations connected with the 1:2 resonance. We  limit the analysis to systems with reflection symmetry with respect to both degrees of freedom: in this case we should more correctly speak of 2:4 resonance, in view of the structure of the resonant Hamiltonian \cite{conto}; however we keep the more standard notion of {\it symmetric 1:2 resonance}. On this ground, although this case retains all the characteristics of a low-order resonance, it also presents some aspects of higher-order resonance systems \cite{sa,tv} and again makes the results of particular relevance for galactic motions. In low-order resonances the bifurcation of new families of periodic orbits is related with loss of stability of one of the normal modes, whereas in higher-order resonances the new families appear in a {\it resonance manifold} from the breakdown of a resonant torus. The peculiarities of the bifurcation sequences from the normal modes are worthy of note by themselves, since the approach followed to study systems with a single symmetry like the spring-pendulum, is not able in the case of double symmetry to unveil the generic behavior of the system \cite{dui}.

In this work we go a little bit in this direction by considering a generic perturbation up to the degree necessary to include resonant terms. Even if this is still not enough to deduce the general behavior for arbitrary perturbations, it allows us to gather complete informations on the bifurcation structure near the resonance of the truncated system. In fact, we will see that a truncation of the normal form at the first term incorporating the resonance is able to capture the essential features of the bifurcation showing how the inclusion of higher order terms in the perturbation is necessary to remove some degeneracies. We apply the procedure of the Lie transform normalization whose algorithmic structure helps in the application of higher-order perturbation approaches. Referring to applications in galactic dynamics, where several numerical and analytical investigations are available \cite{mes,scu,pl1}, we briefly present results concerning a class of systems with elliptical equipotentials including the well known cored logarithmic potential \cite{Bi,MP2}.

The plane of the paper is as follows: in section 2 we introduce the procedure to construct the approximating integrable system by recalling the method of the {\it Lie transform} \cite{gior}; in sections 3 we apply this approach to investigate general aspects of the dynamics obtaining second-order estimates of the bifurcation thresholds of the 1:2 periodic orbits; in section 4 we analyze the stability of normal modes and periodic orbits in generic position; in section 5 we apply these general result to the case of systems with elliptical equipotentials and finally in section 6 we discuss open problems and possible further developments.

\section{The model and its normal form}
Suppose the system under investigation is given by a natural Hamiltonian

\begin{equation} \label{Hamiltonian}
\mathcal H(x,y,p_x,p_y)=\frac12(p^2_x+p_y^2) + \mathcal{V}(x,y)
\end{equation}
where $\mathcal V$ is a smooth potential with an absolute minimum in the origin and
symmetric under reflection with respect to both  coordinate axes.
We assume the potential to be expanded as a truncated power series

\begin{equation}\label{seriesp}
\mathcal V(x,y)\equiv \sum_{n=0}^N V_{n}(x,y)
\end{equation}
where
$V_{n}$ is a homogeneous polynomial of degree $n+2$.   The truncation order $N$ is determined by the problem under study. In force of the reflection symmetries, the `zero' order term can be written as

\be
V_0=\frac12(\omega_1^2x^2+\omega_2^2y^2)
\ee
and the odd order terms are all zero. The two coefficients of the quadratic term  are written so to represent the linearised harmonic frequencies. A system with a potential of type (\ref{seriesp}) can be treated in a perturbative way as a non-linear oscillator system and the two frequencies of the unperturbed system are precisely given by $\omega_1$ and $\omega_2$. In order to put the system in a form suitable for a perturbative approach, we perform the scaling \cite{dui}

\be
(x,y,p_x,p_y)\rightarrow \e^{-1}(x,y,p_x,p_y),\;\;\;\;\e>0 \label{scaling_tr}
\ee
and also rescale the Hamiltonian (\ref{Hamiltonian}) according to

\be
\mathcal H= \e^2 \widetilde{\mathcal H}. \label{scaling_h}
\ee
Thus, we obtain

\be\label{series_h}
\widetilde{\mathcal H}(x,y,p_x,p_y)=\frac12(p^2_x+p_y^2) + \frac12(\omega_1^2 x^2+\omega_2^2 y^2)+\sum_{j=1}^{N/2} \e^{2j} V_{2j}(x,y).
\ee
In this way the terms of the expansions are ordered in powers of the small perturbation parameter.

In general, neither the original system (\ref{Hamiltonian}) nor its expansion  (\ref{series_h}) are integrable. However, in several cases the dynamics around the equilibrium are {\it regular}, namely the measure of chaotic orbits is exponentially small in the perturbation and 
the features are similar to those of an integrable system in a large fraction of phase space. Therefore we proceed to construct a `normal form'~\cite{bp} for the system, namely a new Hamiltonian series which, in the case of 2 degrees of freedom, is an integrable approximation of the original one. Its structure is suitable to capture the most relevant orbital features of the system.

In particular, the normal form is `non-resonant' when the two harmonic frequencies $\omega_1$ and $\omega_2$ are two real numbers so that their ratio is not rational: in this case the normalization produces a `Birkhoff' Hamiltonian depending only on actions. Therefore, since the new Hamiltonian do not depend on angle variables, we get explicit formulas for actions and frequencies of the {\it box orbits} parented by the $x$- and $y$-axis periodic orbits (the `normal modes')

We can instead assemble a `resonant' normal form by assuming from the start a rational value for the ratio of the harmonic frequencies: this assumption produces the presence in the new Hamiltonian of {\it resonant} terms, namely terms depending on a linear combination of angles with integer coefficients. This expedient is legitimate because, even if the unperturbed system is non-resonant for the real value of the  frequency ratio

\be \label{ratio} \rho = \omega_1/\omega_2, \ee
the non-linear interaction between the degrees of freedom, induced by the coupling terms of the perturbation, produces a resonant value of $\rho$. Its commensurability ratio, say $m/n$ with $ m,n \in {\mathbb{N}} $, is determined by the local ratio of oscillations in the two degrees of freedom. This in turn is responsible for the birth of new orbit families bifurcating from the normal modes or from short-periodic orbits generated by the lower-order resonances.
The trick is then to assume that our system is such that the ratio (\ref{ratio}) is not far from a rational value and then to approximate it by introducing a small `detuning' $\delta$ so that

\be
 \rho = m/n + \delta\label{DET}.
\ee
Afterwards we proceed like if the unperturbed harmonic part would be in exact $m$:$n$ resonance by treating the  remaining part as a higher order perturbation. We speak of a {\it detuned} $m$:$n$ {\it resonance}.

Let us proceed with the case of the $m=1$, $n=2$ resonance, so that (\ref{DET}) translates into

\be\label{DET12}
\delta \doteq \frac{\omega_1}{\omega_2} - \frac12\ee
keeping in mind we are in presence of reflection symmetries about both axes: we will shortly see that, in this case, the normalization procedure must be pushed at least to the fourth degree in $\e$ \cite{bbp}.
To give the system a structure suitable to apply the normalization procedure,
we perform the transformation

\be\label{scal_cc}
x_1 =\sqrt{\omega_1} x, \; \; p_1=\frac{p_x}{\sqrt{\omega_1}}, \; \;
x_2 =\sqrt{\omega_2} y, \; \; p_2 = \frac{p_y}{\sqrt{\omega_2}}
\ee
and introduce the complex variables

\begin{equation}
\left \{ \begin{array}{ll}
z_1=&p_1+i x_1,\;\;\;w_1=p_1-i x_1,\\
z_2=&p_2+i x_2,\;\;\;w_2=p_2-i x_2.
\end{array}\right.
\end{equation}
Since we include  the detuning term in the perturbation, we introduce a rescaled detuning parameter $\tilde\d$ such that $\d=\tilde\d\e^2$. Thus, by redefining the Hamiltonian according to the scaling

\begin{equation}\label{scal_det}
H \doteq 2 \rho \widetilde{\mathcal H}=\left(1 +2 \d\right)\widetilde{\mathcal H}
\end{equation}
 and collecting terms in $\e$,
we put the rescaled Hamiltonian into the form

\begin{equation}
\label{Has}
 H(\vec{z},\vec{w};\tilde\d)=\sum_{j=0}^{N/2} \e^{2j} H_{2j}(\vec{z},\vec{w};\tilde\d)
\end{equation}
where the unperturbed term is given by

\be
H_0(\vec{z},\vec{w})=\frac 12 w_1z_1+ w_2 z_2
\ee
and the detuning term

\be \e^{2} \tilde\d(x_1^2+p_1^2) = \e^{2} \tilde\d  w_1z_1,\ee
rather than being treated as a term of order zero, is considered as a term of order 2.

The system is now ready for a standard resonant normalization (see \cite{bp,SV}). Here we are going to apply the method based on the Lie transform \cite{bp}. An account of the procedure has been given in \cite{MP}; we briefly recall here the main ideas to adapt it to the symmetric case. The starting point is to treat the canonical transformation as a `flow' along the Hamiltonian vector field associated to a generating function. Let us consider a phase-space function expanded as a power series in the canonical variables

\be\label{gene}
G=\sum_{l=1}^{N/2} \e^{2l} G_{2l}
\ee
where the odd-order terms are zero in force of the reflection symmetries.
Since we want the method to work with the series expansion given in (\ref{Has}),
we assume that the non-vanishing terms are polynomials of the form

\be
G_{2l}(\vec{z},\vec{w};\tilde\d)=\sum_{j=0}^{l}\tilde\d^j g_{2(l-j+1)}(\vec{z},\vec{w}),\;\;\; g_m\in \mathcal P_m,\;\;m=0,\dots l.
\ee
where $\mathcal P_m$ denotes the space of homogenous polynomials of degree $m+2$
 in the $(\vec{z},\vec{w})$ coordinates which are invariant with respect to the reflections

\ba\label{simm}
(z_1,z_2,w_1,w_2)&\rightarrow&(-z_1,z_2,-w_1,w_2) \\
(z_1,z_2,w_1,w_2)&\rightarrow&(z_1,-z_2,w_1,-w_2)
\ea
and their combinations. To the series (\ref{gene}) is naturally associated the linear differential operator
\be\label{diff}
\rm e^{\mathcal L_G} = \sum_k \frac1{k!} {\cal L}_G^k,\ee
whose action on a generic function $F$ is given by the Poisson bracket:
\be
{\cal L}_G F \doteq \{F,G\}.\ee

The original Hamiltonian system (\ref{Has}) undergoes a canonical transformation to new variables $
(\vec{Z},\vec{W}),$ such that the new Hamiltonian is

\begin{equation}\label{HK}
     K(\vec{Z},\vec{W})={\rm e}^{{\cal L}_G} H(\vec{z},\vec{w}),
\end{equation}
where $K$ is assumed in the form of a series expansion similar to  (\ref{Has}), namely

\be\label{NF}
K(\vec{Z},\vec{W})=\sum_{j=0}^{N/2}\e^{2j}K_{2j} (\vec{Z},\vec{W}).
\ee
To construct $K$ starting from  (\ref{Has}) is a recursive procedure exploiting an algorithm based on the Lie transform \cite{bp,gior}. To understand how it works, let us consider the first step of the procedure, namely let us perform a transformation given by a function $G_2$ considered as the first term in the generating function (\ref{gene}). The general relation (\ref{HK}) takes the form

\be\label{HK1}
K_0 + \e^2 K_2 + ... = (1 + {\cal L}_{G_2} + ...) (H_0 + \e^2 H_2 + ...).
\ee
By equating polynomials of the same order in $\e$, we get the system:

\begin{eqnarray}
K_0&=&H_0 \label{eq_ord0}\\
K_2&=&H_2-\mathcal L_{G_2}H_0\label{eq_ord1}\\
K_4&=&H_4-\mathcal L_{G_2}H_2-\frac12\mathcal L^2_{G_2}H_0\\
&\dots\dots&\nonumber\\
K_n&=&\mathcal L_{G_2}H_0+R_n\label{eq_ord_n}
\end{eqnarray}
where the `rest' $R_n$ contains terms which are known if the preceding $n-1$ equations have been solved.

Equation (\ref{eq_ord0}) simply states that the zero order new Hamiltonian coincides with the zero order
(unperturbed) one. To proceed, we have  to solve the second equation to find $K_2$, a  differential equation involving {\it two} unknown functions, $K_2$ and $G_2$. To overcome this problem we have to make some decision about the structure the new Hamiltonian $K$ must have, that is we have to choose  a \emph{normal form} for it. We make the choice that $K$ has to satisfy

\begin{equation}
\{K,H_0\}=0
\end{equation}
or equivalently

\begin{equation}
\mathcal L_{H_0}K=0. \label{norm_condition}
\end{equation}
In this way the system with Hamiltonian $K$ admits $H_0$ as a new integral of motion. To satisfy condition (\ref{norm_condition}), let us consider for greater generality, the  $n$-th step in the normalization procedure and impose that $K_n$ and $G_n$  be solutions of the system

\begin{equation}
\left\{\begin{array}{ll}
         \mathcal L_{G_n}H_0+R_n & = K_n \\
         \mathcal L_{H_0}K_n & =0.
       \end{array}
       \right.
\end{equation}
We rewrite the first equation as

\begin{equation}\label{homological_eq}
\mathcal L_{H_0}G_n+K_n=R_n.
\end{equation}
This is the so called \emph{homological equation}. By the way, in force of the reflection symmetry, for
odd order terms equation (\ref{homological_eq}) gives the trivial solution $K_n=G_n=0$, proving the consistency of the assumptions concerning the expansions (\ref{gene}) and (\ref{NF}). For $n=2l$, if we
recall that the detuning parameter is  assumed of order two, equation (\ref{homological_eq}) becomes

\be\label{omo_s}
\sum_{j=0}^{l}\tilde\d^j \mathcal L_{H_0}g_{2(l-j+1)}+ \sum_{j=0}^{l}\tilde\d^jk_{2(l-j+1)}=\sum_{j=0}^{l}\tilde\d^jr_{2(l-j+1)},\;\;\;
\ee
with $g_m,k_m$ and $r_m \in\mathcal P_m$, $m=0,\dots l$. Since $\mathcal L_{H_0}:\mathcal P_m\rightarrow$ $\mathcal P_m$,
to solve equation (\ref{omo_s}) is equivalent to solving $l=n/2$ equations of the type

\be\label{omo_eqns}
\mathcal L_{H_0}g_{2(l-j+1)}+k_{2(l-j+1)}=r_{2(l-j+1)},\;\;\; j=0,\dots,\frac{n}{2}.
\ee
Now, thanks to the semisimple character of the linear operator $\mathcal L_{H_0}$, its kernel and its range are in direct sum over the space $\mathcal P_m$. This implies that the system of equations (\ref{omo_eqns}) and hence the $n$-th homological equation, can always be solved if $K_n$ satisfies (\ref{norm_condition}).

We have so far constructed the normal form $K_0 + \dots + K_n$ and the generating function  $G_2 + \dots + G_n$: we use them to compute $K_{n+2}$ and $G_{n+2}$ and so for. However, as we have observed above, the series are divergent, thus we must truncate the procedure at some finite
order, say $M$. In the case of a $m=1$, $n=2$ resonance in the presence of reflection symmetries about both axes, the normalization procedure must be pushed at least to order $M=4$. Therefore, generalizing the assumptions made in \cite{tv}, we assume that the non-vanishing terms in the series expansion of the original Hamiltonian are given by

\ba
H_0&=&\frac12(x_1^2+p_1^2)+(x_2^2+p_2^2) \label{hp0}\\
H_2&=&\tilde\d(x_1^2+p_1^2)-\frac14 (a x_1^4+ b x_2^4+ c x_1^2 x_2^2)  \label{hp2}\\
H_4&=&\frac16 \left(a_1 x_1^6+b_1 x_1^4 x_2^2+c_1 x_1^2 x_2^4+d_1 x_2^6\right),  \label{hp4}
\ea
where the arbitrary coefficients  appearing in the higher order terms represent the most general potential truncated to degree six in the coordinates and complying with the enforced double reflection symmetry. The choice of coefficients and signs is suggested by the values the coefficients take in the commoner physical cases: we remark that, in applying the results obtained in the following to specific model problems, we have to take into account that the Hamiltonian (\ref{hp0}--\ref{hp4}) is in the form `prepared' for normalization. Therefore the canonical variables are rescaled according to (\ref{scal_cc}) and the frequency ratio is expanded in series of the detuning as in (\ref{DET12}). These transformations affect the numerical values of the coefficients of the various terms.

The outcome of the normalization procedure described above is a series of the form (\ref{NF}) that, by introducing  action-angle like variables
by means of

\ba
\left\{\begin{array}{ll}
\vec{Z}=&\;i\sqrt{2 \vec{J}}\  {\rm e}^{-i \vec{\theta}}\\
\vec{W}=&-i\sqrt{2 \vec{J}}\  {\rm e}^{i \vec{\theta}},
\end{array}\right.
\ea
can be written as

\be\label{NF4}
K(\vec{J},\vec{\theta})=\sum_{j=0}^{2}\e^{2j}K_{2j}(\vec{J},\vec{\theta}),
\ee
which, coherently with the truncation order $M=4$, implies that a reminder of order six in $\e$ is neglected. In these canonical variables it is possible to express the series in a manageable form whereas the use of Cartesian variables (both in real or complex form) gives quite cumbersome formulas. The non-vanishing terms in the normal form turn out to be the following:

\ba
K_0 &=& J_1+2J_2 \\
K_2 &=& 2\tilde\d J_1-\frac38 a J_1^2 - \frac14 c J_1 J_2 - \frac 38 b J_2^2\\
K_4 &=& \left(-\frac{17}{64} a^2 + \frac{5}{12} a_1\right) J_1^3 +
               \left(-\frac{17}{128} b^2  + \frac{5}{12} d_1   \right) J_2^3  + \frac{1}{192}
               \left(-18bc - 5 c^2 + 48 c_1 \right) J_1 J_2^2  \nn
        && + \frac{1}{192} J_1^2 J_2 \left[-\left(36a + 5c \right) c + 48 b_1 + \left(3(a-c)c+8 b_1 \right) \cos(4 \theta_1-2 \theta_2)\right] .
\ea
The resulting Hamiltonian (\ref{NF4})
is the basis for all subsequent work. We observe that up to order 2, the dependence is only on action-like variables: therefore, up to this order they are true conserved actions with angles evolving linearly on the invariant tori. The resonant combination of the angles with zero phase difference appears only in the term of order 4 (and higher) and this is the reason why this is the minimum order required in this study. In ref.\cite{tv} it is proven that this phase value always occur in potential problems and that in more general Hamiltonian systems the phase can be different from $0,\pm\pi$.

The resonant terms determine a nonlinear coupling of the two degrees of freedom. Although the system is integrable by construction, the solution of the canonical equations is in general not expressible in terms of elementary functions. We can however exploit the second conserved quantity to reduce the system and we show in the subsequent sections how to use this technique to gain understanding of the structure of the phase-space.

\section{Bifurcation analysis of the 1:2 symmetric resonance}

The essential information we need concerns the existence and stability of the periodic orbits associated to the resonance \cite{bbp,MP}. We can compute the thresholds for the bifurcations sequences in terms of the parameters relying on the regular nature of the dynamics given by the normal form.
In two degrees of freedom, if a Hamiltonian possesses a second independent integral of motion, the system is {\it Liouville integrable}.
Due to the normalization procedure, we have obtained the Hamiltonian (\ref{NF4}) with the second independent integral of motion

\begin{equation}\label{ene12}
K_0=J_1+2J_2.
\end{equation}
We can use this integral to reduce the dimension of the problem by performing the canonical transformation to
`adapted resonance coordinates' \cite{SV}

\begin{equation}\label{calr12}
\left\{\begin{array}{ll}
  J_1=&\E+2\R  \\
   J_2=& 2\E-\R \\
\psi=&4\theta_{1}- 2\theta_{2} \\
\chi=&2\theta_{1}+ 4\theta_{2}
\end{array}\right.
\end{equation}
It can be easily seen that $\chi$ is cyclic and its conjugate momentum is proportional to the additional integral of motion, namely

\be
\E=(J_1+2J_2)/5.
\ee
The value of $\E$ plays a special role in the discussion of the orbit structure and we will refer to it as the {\it distinguished} parameter \cite{Br2:1}. Thus we introduce the `reduced Hamiltonian'

\be\label{Ker12}
\mathcal K(\mathcal R, \psi; \E) = 5\E + \e^2\mathcal K_1(\mathcal R)
        + \e^4 \mathcal K_2(\mathcal R,\psi).
\ee
In the computation of \eqref{Ker12} and through the rest of the paper the use of algebraic manipulators
is almost indispensable. By using {\sc mathematica} \textregistered \ we obtain the following expressions
for $\mathcal K_1$ and $\mathcal K_2$:

\ba
\mathcal K_1(\R)&=&-\frac18 \left[3b(-2\E + \R)^2 + 3 a(\E + 2\R)^2
                     + 2(\E + 2 \R)(2 c\E - c \R - 8\tilde\d)\right] \\
\mathcal K_2(\R,\psi)&=&\left(-\frac{17}{64} (a^2 + 4 b^2)-\frac{3}{8} c (a+ b)-\frac{5}{32} c^2 +\frac{5 }{12} a_1+\frac{1}{2} b_1 + c_1+\frac{10 }{3}  d_1\right)\E^3+\mathcal A (\R)+\mathcal B(\R)\cos\psi
\ea
with

\ba
\mathcal A(\R)&=&\frac{\R^3}{384} \left(-816 a^2+51 b^2+288 a c-72 b c+20 c^2+1280 a_1-384 b_1+192 c_1\right. \nn
&-&\left. 160 d_1\right)+\frac{\E \R^2}{64}  \left(-204 a^2-51 b^2-48 a c+42 b c+5 c^2+320 a_1+64 b_1\right.\nn
&-&\left.112 c_1+160 d_1\right)-\frac{\E^2 \R }{192} \left(306 a^2-306 b^2+252 a c+72 b c+55 c^2\right.\nn
&-&480 a_1\left.-336 b_1-192 c_1+960 d_1\right),\\
\mathcal B(\R)&=&\frac{1}{192} (2 \E-\R) (\E+2 \R)^2 \left(3ac-3c^2 +8 b_1\right).
\ea
Considering the dynamics at a fixed values of $\E$, we have that $\mathcal K$ defines a one degree of freedom system with the following equations of motion

\ba
\dot\psi&=&\frac14\left(6 b\E - 3 c\E - 3b \R + 4 c \R -
        6 a(\E+ 2\R)+ 16\tilde\delta\right)\e^2 + \left( \frac{\partial \mathcal A}{\partial \R}+ \frac{\partial \mathcal B}{\partial \R}\cos\psi\right)\e^4 \label{psipunto12}\\
\dot\R &=&\frac{\mu}{192}  \left((2 \E-\R) (\E+2 \R)^2  \sin\psi\right) \e^4 \label{rpunto12}
\ea
where

\be\label{mu}
\mu \doteq 3 a c-3 c^2+8 b_1.
\ee
The fixed points of this system give the periodic orbits of the original system.

The pair of fixed points with ${\cal R}=2\E$, $\R=-\E / 2$ correspond to the  \emph{normal modes}, that is to the periodic orbit along the $x$-axis $(J_2 = 0)$ and to the periodic orbit along the $y$-axis  $(J_1 = 0)$ respectively. Additional periodic orbits may appear when the system passes through the resonance. These periodic orbits `in general position' exist only above a given threshold in the distinguished parameter $\E$ when the axial orbits change their stability. This phenomenon can be seen as a {\it bifurcation} of the new family from the normal mode when it enters in 1:2 resonance with a normal perturbation. The phase between the two oscillations determines the nature of the families: they are respectively given by the conditions $\psi = 0$ ({\it banana} orbits) and $\psi = \pm \pi$ ({\it anti-banana} orbits), where we use the nicknames introduced in ref.\cite{mes}. These phase conditions are solutions of ${\dot {\cal R}}=0$ (when ${\cal R}\neq2\E$ and  $\R\neq-\E / 2$) and determine the corresponding solutions of ${\dot \psi} = 0$. 

Let us start looking for banana orbits. Following a standard approach \cite{henrard}, we set $\psi=0$ in the equation $\dot\psi=0$ and look for a solution in the form

\be\label{Rtb}
\R=\R_0+\R_1\e^2+O(\e^4)
\ee
so that the righthand side of \eqref{psipunto12} vanish up to fourth order in $\e$.
 For $\psi=0$, we substitute (\ref{Rtb}) in (\ref{psipunto12}) and collect terms of the same order in $\e$. Equating to zero the coefficient of the second order term in $\e$, we find that $\R_0$ has to satisfy

\be
3(2b-2a-c)\E+(4c-3b-12a)\R_0+16\tilde\d=0
\ee
This equation admits solution only if

\be\label{prima_cond}
\nu \doteq 12a+3b-4c\neq0
\ee
and, if this condition is satisfied, we find

\be
\R_0=\R_{B0}\doteq\frac{-3(2 a -2 b+ c )\E+16 \tilde\delta }{12 a+3 b-4 c}.
\ee
Once $\R_0$ is computed, we look for $\R_1$ such that the fourth order term of the righthand side
of \eqref{psipunto12} vanish. Since (\ref{prima_cond}) is satisfied, we find  a solution

\be
\R_1=\R_{B1}(\E;\d).
\ee
The corresponding fixed point is given by

\be
\R=\R_{B}=\R_{B0}+\R_{B1}\e^2,\;\;\;\psi=0
\ee
 and determines the banana orbits (there are \emph{two} of them):

\ba
J_1&=&J_{1B}=\E+2\R_{B} \\
J_2&=&J_{2B}=2\E-\R_{B}.
\ea
Similarly, if (\ref{prima_cond}) is satisfied, for $4\theta_{1}- 2\theta_{2}=\pm\pi$, we find
\ba
J_1&=&J_{1A}=\E+2\R_{A} \\
J_2&=&J_{2A}=2\E-\R_{A},
\ea
which correspond to the anti-banana orbits. In view of (\ref{calr12}), the constraints

\be
 0 \le J_1\leq5\E,\;\;\;\;\; 0\leq J_2\leq\frac{5 {\cal E}}{2},\label{ex_cond}
 \ee
applied to these solutions give the condition of existence for these periodic orbits in general position.
Whether these conditions are satisfied or not, it depends on the parameters of the system. For the banana
orbits we find that at the zero perturbative order

\ba
J_{1B}=\frac{5(3b-2c)\E+32\tilde\d}{12a+3b-4c}, \nn
J_{2B}=\frac{5(6a-c)\E-16\tilde\d}{12a+3b-4c}.
\ea
Thus, we get different existence conditions according to the sign of the constant $\nu$ defined in (\ref{prima_cond}).
Namely, taking $\e$ small enough such that the constraints (\ref{ex_cond}) remains satisfied up to the first perturbative order, banana orbits bifurcate in the following cases:
\ba
 \hbox{if } \tilde\d\nu>0 \hbox{ and }\tilde\d (6a-c)>0 &,&
                                   \left\{ \begin{array}{ll}
                                      \hbox{if }\tilde\d (3b-2c)>0: & \E>\E_{B1} \\
                                     \hbox{if } \tilde\d (3b-2c)<0: & \E_{B1}<\E<\E_{B2} \\
                                   \end{array} \right.,  \label{bxp} \\
                                   \hbox{if } \tilde\d\nu<0 \hbox { and } \tilde\d (3b-2c) < 0&,&  \left\{ \begin{array}{ll}
                                      \hbox{if } \tilde\d (6a-c)<0: & \E>\E_{B2} \\
                                     \hbox{if } \tilde\d (6a-c)>0: & \E_{B2}<\E<\E_{B1} \\
                                  \end{array} \right., \label{byp}
 \ea
where the critical values

\ba
{\cal E}_{B1}&\doteq&\frac{16  }{5(6 a-c)}\tilde\d-\frac{16  \left(306 a^2-33 a c-8 c^2-480 a_1+56 b_1\right)}{15 (6 a-c)^3}\tilde\d^2\e^2+O(\e^4)
\label{EPOB1}\\
\E_{B2}&\doteq&\frac{32 }{5 (2 c-3b)} \tilde\delta-\frac{16 \left(153 b^2-72 b c-20 c^2+192 c_1-480 d_1\right)}{15 (3 b-2 c)^3} \tilde\delta ^2\e^2+ O(\e^4) \label{EPOB2}
\ea
correspond to the solutions of $J_{2B}=0$ and $J_{1B}=0$ and respectively determine the bifurcation of the banana orbits from the $x-$normal mode and the $y-$normal mode in the first sub-cases of (\ref{bxp}) and (\ref{byp}). In the other two sub-cases 
the periodic orbit bifurcating from one of the normal modes disappears on the other.

A similar argument provides the existence condition of anti-banana orbits. Since to the first perturbative
order $J_{1A}=J_{1B}$ and $J_{2A}=J_{2B}$, the birth of anti-bananas is given by the same conditions on the coefficients given above. However the higher order terms in $J_{kA}$, $J_{kB}$, $k=1,2$ are in general different: the discrimination between the thresholds for the bifurcation of the two families is possible only by going at least to second order \cite{bbp}. This result was expected on the basis of the structure of the new Hamiltonian (\ref{NF4}).

Nevertheless, up to the second order in $\e$ we find that the threshold value corresponding to the bifurcation of anti-banana orbits from the $y-$axis, $\E_{A2}$, coincides with the critical value $\E_{B2}$ of (\ref{EPOB2}). Hence, {\it if banana and anti-banana orbits bifurcate from the $y-$normal mode, they do it concurrently}. On the other hand the bifurcation from the $x-$axis orbit occurs at

\be
\E=\E_{A1}, \; \; \; \; \; \;   \tilde\d (6a-c)>0,
\ee
where

\be
\E_{A1}\doteq \frac{16}{5(6 a-c)}\tilde\delta-\frac{16\left(306 a^2-39 a c-2 c^2-480 a_1+40 b_1\right)}{15 (6 a-c)^3}\tilde\d^2\e^2 + O(\e^4) \label{EPOA1}
 \ee
which, at second order, is different from (\ref{EPOB1}). Thus, we obtain the following existence conditions for anti-banana orbits:

\ba
 \hbox{if } \tilde\d\nu>0 \hbox{ and } \tilde\d (6a-c)>0 &,&
                                   \left\{ \begin{array}{ll}
                                      \hbox{if } \tilde\d (3b-2c)>0: & \E>\E_{A1} \\
                                     \hbox{if }  \tilde\d (3b-2c)<0: & \E_{A1}<\E<\E_{A2}\equiv\E_{B2} \\
                                   \end{array} \right. , 
 \label{axp}\\
 \hbox{if } \tilde\d\nu<0 \hbox { and } \tilde\d (3b-2c)<0&,&  \left\{ \begin{array}{ll}
                                      \hbox{if } \tilde\d (6a-c)<0: & \E>\E_{B2} \\
                                     \hbox{if }  \tilde\d (6a-c)>0: & \E_{B2}<\E<\E_{A1} \\
                                  \end{array} \right. , \label{ayp}\\
\hbox{with } && \left\{
                                    \begin{array}{ll}
                                      \E_{A1}\geq\E_{B1}, & \hbox{ if }\tilde\d\mu>0 \\
                                      \E_{A1}<\E_{B1}, & \hbox{ if }\tilde\d\mu<0
                                    \end{array}
                                  \right. .
\ea

We can write explicitly the relative magnitude of the two thresholds for bifurcation from the normal modes because we find

\ba
\E_{A1}-\E_{B1} &=& \frac{32}{15} \frac{\mu}{(6a-c)^3}\tilde\d^2\e^2 + O(\e^4),\label{GAB1}\\
\E_{A2}-\E_{B2} &=& 0 + O(\e^4).\label{GAB2}\ea
The first of these expressions points out that the hierarchy of bifurcations from the $x-$normal mode is determined by the sign of the constant $\mu$ defined in (\ref{mu}). The second confirms that, at the order of our perturbative treatment, the bifurcations from the $y-$normal mode occur simultaneously. Which of the two scenarios is actually happening depend on the value of the parameters according to the conditions listed above.

\begin{figure}[h!]
\centering
\includegraphics[width=0.49\columnwidth]{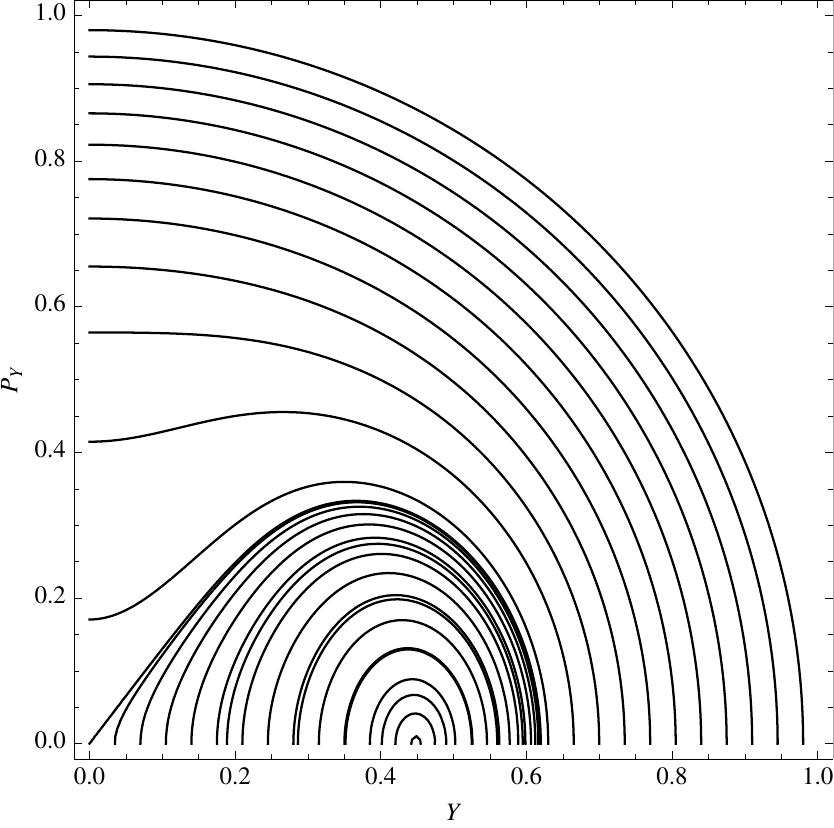}
\includegraphics[width=0.49\columnwidth]{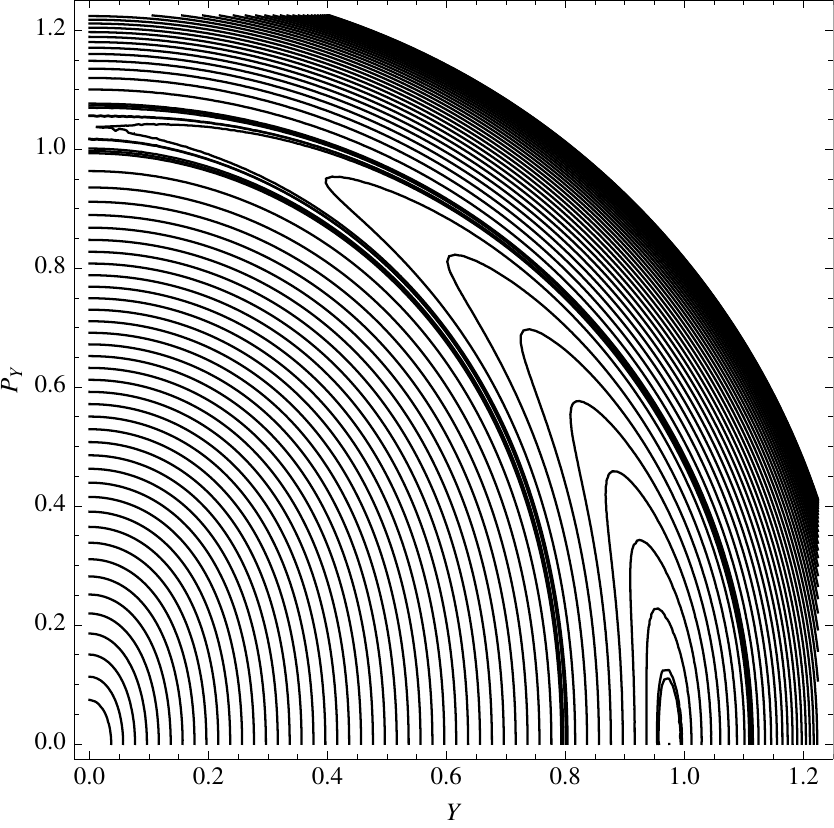}
\includegraphics[width=0.49\columnwidth]{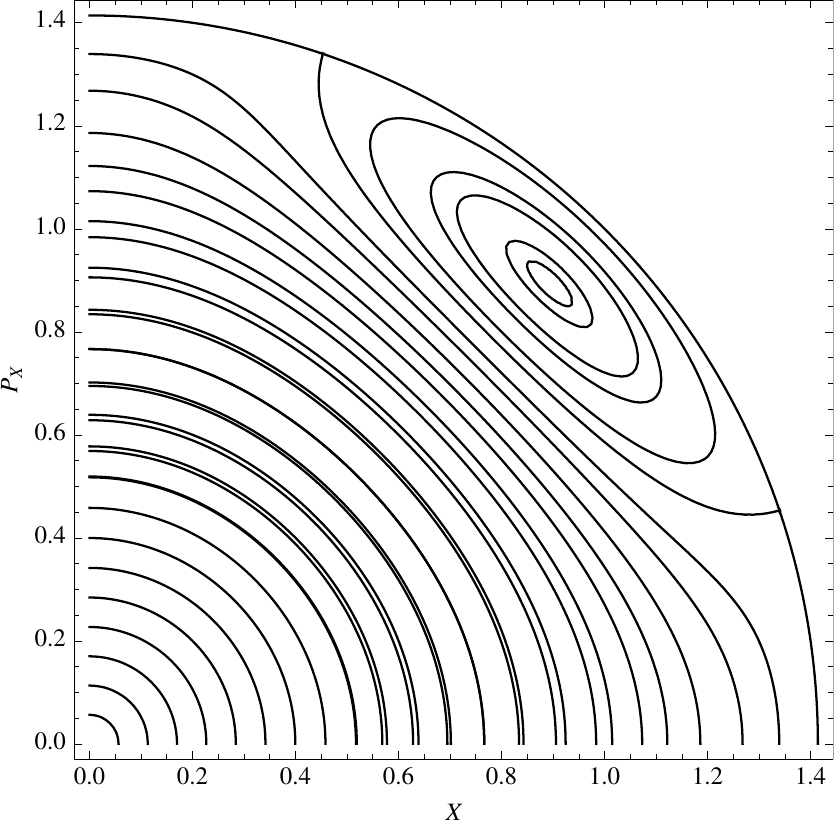}
\includegraphics[width=0.49\columnwidth]{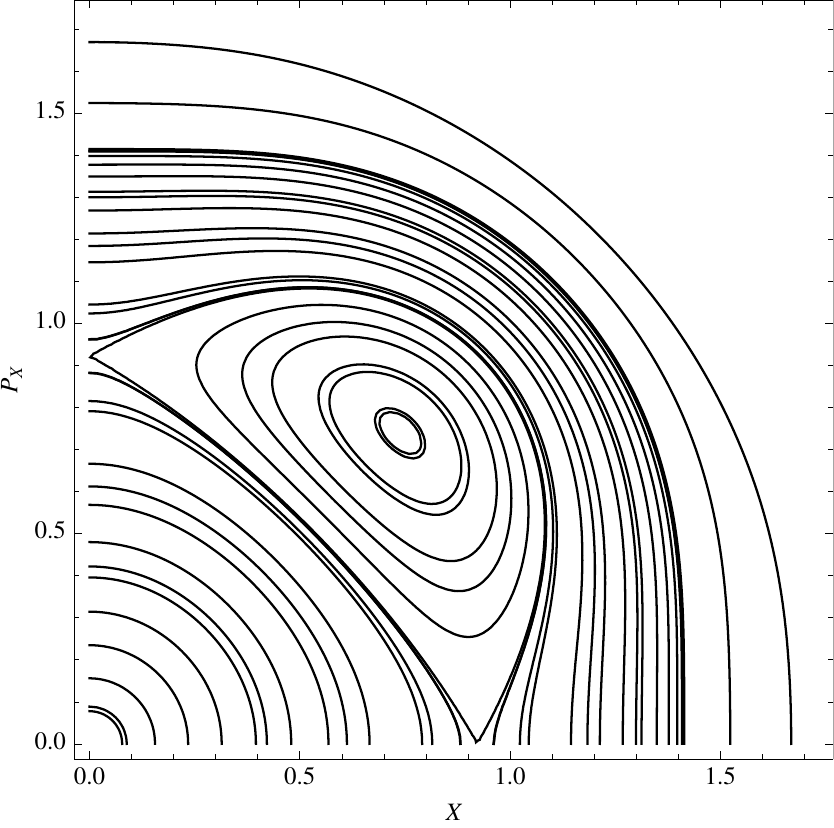}

\caption{Dynamics of the 2:4 doubly-symmetric resonant Hamiltonian, see text for explanation.}
\label{pss}
\end{figure}

In Fig.\ref{pss} we plot the surfaces of sections relevant for the first scenario concerning the bifurcations from the $x-$normal mode: the upper panels are the $(x,p_x)$ sections after the bifurcation of the stable banana (left) and after the subsequent bifurcation of the unstable anti-banana (right); the lower panels are the corresponding $(y,p_y)$ sections. This occurrence is typical of a lower order bifurcation: the normal mode losing stability at the first bifurcation is the origin in the upper panels and is the last contour in the lower. The contour in the lower left panel is the unstable mode that regains stability in the right panel.

The second scenario in which the two families appear together from the $y-$normal mode is the typical occurrence in higher-order resonances: pictorially it correspond to a transition from the non-resonant invariant tori around the normal mode {\it directly} to the appearance of the resonance manifold as it emerges in the two lower panels with stable and unstable orbits arising in pair.

For practical purposes, e.g. to compare these findings with the outcomes of numerical simulations, it would be more useful to have the expression of the bifurcation curves in terms of `physical' parameters. The most natural way to represent the thresholds is that plotting curves in the $(E,\d)-$plane, where $E$ is the physical energy of the system defined by (\ref{Hamiltonian})

\begin{equation}
\mathcal H(x,y,p_x,p_y)=E
\end{equation}
and $\d$ is the `true' detuning defined in (\ref{DET}). According to the rescaling (\ref{scal_det}), on the $x-$axis orbit ($J_2=0$, $J_1=5\E$), we have

\be \label{ks}
\e^2 K= 5\E\e^2 +\left(- \frac{75 a}{8} \E^2 +
          10\E\tilde\d \right)\e^4+\dots=(1+2\tilde\d\e^2) E.
\ee
The dots are present to recall that a reminder has been neglected. The series from equation (\ref{ks}) can be
used to express the physical energy $E$ in terms of $\E$ \cite{bbp,pbb}, namely

\be
E=5 \E\e^2 -\frac{75}{8}a\E^2\e^4 + O(\e^6).
\ee
Thus, up to the second order in $\e$, for $\E$ satisfying equations (\ref{EPOB1}) and (\ref{EPOA1}) and $\d$ as in \eqref{DET12},  we obtain the following threshold values

\ba
E_{B1}&=&\frac{16 }{6 a-c}\delta-\frac{16 \left(414 a^2-51 a c-8 c^2-480 a_1+56 b_1\right)}{3 (6 a-c)^3} \delta ^2  +O(\d^3), \label{E1B}
\\
E_{A1}&=&\frac{16 }{6 a-c}\d-\frac{16 \left(414 a^2-57 a c-2 c^2-480 a_1+40 b_1\right)}{3 (6 a-c)^3}\d^2+O(\d^3), \label{E1A}
\ea
for the bifurcation of respectively banana and anti-banana orbits from the  $x-$axis orbits.

A similar argument gives the threshold value for the bifurcations from the $y-$axis orbit. Since $\E_{A2}=\E_{B2}$, we have
$E_{A2}=E_{B2}$ and by using the relation between the true energy and the distinguished parameter on the normal mode we get

\ba
E_{B2}=\frac{32 }{2 c - 3 b}\d-\frac{16 \left(99 b^2+36 b c-68 c^2+192 c_1-480 d_1\right)}{3 (3 b-2 c)^3} \delta ^2+O(\d^3). \label{E2B}
\ea

\section{Stability analysis of the 1:2 symmetric resonance}\label{stability}

Let us now consider the question of the stability of periodic orbits: this analysis complements that of the previous section allowing us to test the relation between change of nature of normal modes and bifurcation of a new family. For banana and anti-banana orbits an ordinary investigation  of the equations of variations of the system is enough to perform the linear stability analysis. However, in the case of axial orbits, action angle variables have singularities on them and this also affects the adapted resonance coordinates. However the remedy to this problem is quite straightforward: to use a mixed combination of action angle variables
on the orbit itself and Cartesian variables for the other dof.

Let us start with the stability analysis of the periodic orbits in general position. We have to investigate the fate of a normal perturbation of the periodic orbit under test. The system of differential equations for the perturbations $(\d\psi,\d\R)$ is given by

\be
\frac{d}{dt}\left(
              \begin{array}{c}
                \d\psi \\
                \d\R \\
              \end{array}
            \right)=\left(
                      \begin{array}{cc}
                        \mathcal K_{\R\psi} & \mathcal K_{\R\R} \\
                       -\mathcal K_{\psi\psi} & -\mathcal K_{\R\psi} \\
                      \end{array}
                    \right)\left(
              \begin{array}{c}
                \d\psi \\
                \d\R \\
              \end{array}
            \right).
\ee
Here we again use the reduced Hamiltonian (\ref{Ker12}) and, with a small abuse of notation, we assume without denoting it explicitly that the entries in the Hessian matrix are evaluated on the specific orbit we are interested in. Then, the sign of the determinant

\be
\Delta(\R,\psi;\E,\tilde\d)=\mathcal K^2_{\R\psi}-\mathcal K_{\R\R}\mathcal K_{\psi\psi}
\ee
computed on the periodic orbit determines the fate of the
perturbation: if $\Delta(\R,\psi;\E,\tilde\d)$ is negative it gives the frequency of bounded oscillating solutions thus determining stability; a change of sign, as a consequence of varying $\E$, produces a stability transition.

On the banana and anti-banana orbits we respectively have

\ba\Delta(\R_B,0)&=&
\Delta_B\doteq\mu\frac{(30 a \E-5 c \E-16 \tilde\delta ) (15 b \E-10 c \E+32 \tilde\delta )^2 }{768 (12 a+3 b-4 c)^2}\e^6+O(\e^8)\label{detb}
\\
\Delta(\R_A,\pi)&=&\Delta_A\doteq-\mu\frac{(30 a \E-5 c \E-16 \tilde\delta ) (15 b \E-10 c \E+32 \tilde\delta )^2 }{768 (12 a+3 b-4 c)^2}\e^6+O(\e^8) \label{deta}
\ea
and thus we see that the parameter  $\mu$ plays an important role also for stability. Comparing with  (\ref{EPOB1}) and (\ref{EPOB2}), for $\tilde\d\mu>0$ banana orbits are stable in the case they bifurcate from the $x$-axis orbit ($\E>\E_{B1}$, if $\tilde\d\nu>0$ and $\tilde\d N>0$) and unstable in case their bifurcation occurs from the $y$-axis orbit. Otherwise, we have instability (stability) when the bifurcation occurs
from the $x$-normal mode ($y$-normal mode).
Since $\Delta_A=-\Delta_B$ up to the third perturbative order, anti-banana orbits turn out to be unstable when banana orbits are stable and viceversa. Actually, the fourth order terms in (\ref{detb}) and (\ref{deta}) are different, but their difference is again a multiple of $\mu$.

We have also seen in (\ref{GAB1}) that the bifurcation order from the $x$-axis depends on the sign of $\tilde\d\mu$. Thus, we can now state that, if $\tilde\d\mu$ is positive, we have at first the bifurcation of (stable) banana orbits followed by (unstable) anti-banana orbits. On the contrary, for negative values of $\tilde\d\mu$ the bifurcation order and stability nature are inverted.

Let us now study the stability of the normal modes. Considering the $x-$axis orbit, we use action-angle variables on the orbit and Cartesian variables on the normal bundle to it, namely

\be
\left\{
  \begin{array}{cc}
    X=& \sqrt{2J}\cos\theta \\
    P_X= &\sqrt{2J}\sin\theta\\
    Y=& Y \\
    P_Y=& V
  \end{array}
\right.
\ee
so that the periodic orbit is given by

\be
Y=V=0,\;\;\;\;\;\; J=5\E.\label{x-ax}
\ee
In these coordinates, the system of differential equation for the perturbations of the normal mode is given by

\be
\frac{d}{dt}\left(
              \begin{array}{c}
                \d Y \\
                \d V \\
              \end{array}
            \right)=\left(
                      \begin{array}{cc}
                        \widetilde {\mathcal K}_{VY} & \widetilde {\mathcal K}_{VV} \\
                       -\widetilde {\mathcal K}_{YY} & -\widetilde {\mathcal K}_{YV} \\
                      \end{array}
                    \right)\left(
              \begin{array}{c}
                \d Y \\
                \d V \\
              \end{array}
            \right)\label{var_x}
\ee
where $\widetilde {\mathcal K}=\mathcal K(Y,V,\theta,J)$. However the matrix of the second derivatives of $\widetilde{\mathcal K}$ on the periodic orbit depends on $\theta(t)=\omega t$, where

\ba
\omega=\frac{\partial\widetilde{\mathcal K}}{\partial J} = 1-\left(\frac{15 a \E }{4}-2 \tilde\delta  \right)\e^2
 -  \left(\frac{1275}{64} a^2 + \frac{15}4 a_1  \right) \E^2 \e^4+O(\e^6).
\ea
To remove the dependence on time we introduce complex coordinates

\be
\left\{
  \begin{array}{ll}
    z= & Y+i V \\
    w= & Y-i V
  \end{array}
\right.\label{compl_x}
\ee
and perform the `rotation'

\be
\left\{
  \begin{array}{ll}
  z = & Z e^{-2i \omega  t}, \\
  w = & W e^{2i \omega  t}.\end{array}
\right.\ee
In this way, the equations of variation (\ref{var_x}) on the periodic orbit \eqref{x-ax} give

\be
\frac{d}{dt}\left(
              \begin{array}{c}
                \d Z \\
                \d W \\
              \end{array}\right)=
           i \left(
                      \begin{array}{cc}
                        \Lambda_{11} & \Lambda_{12} \\
                       \Lambda_{21} & \Lambda_{22} \\
                      \end{array}
                    \right)
             \left( \begin{array}{c}
                \d Z \\
                \d W \\
              \end{array}
            \right)
\ee
where, up to the second perturbative order,

\ba
\Lambda_{11}&=&-\Lambda_{22}=\frac{1}{4} (30 a \E-5 c \E-16 \tilde\delta ) \e^2
 + \frac{25}{192}  \left(306 a^2 -36 a c -5 c^2 -480  a_1+48  b_1\right)\E^2 \e^4\\
\Lambda_{12}&=&-\Lambda_{21}=-\frac{25}{192} \mu\E^2 \e^4.
\ea
By solving $\det \Lambda=0$ we find, as expected, that the critical values of $\E$ which determines a change in the stability of the $x-$axis orbit are precisely given by the bifurcation values

\be
\E=\E_{B1}\;\;\;\hbox{ and }\;\;\;\E=\E_{A1}
\ee
as defined by  (\ref{EPOB1}) and (\ref{EPOA1}). Regardless of the the nature of the occurring bifurcations (this is given by the sign of $\mu$), the first one produces a transition from stability to instability of the $x-$normal mode and the second one a return to stability.

Let us now consider the stability of the $y-$axis orbit. Since the periodic orbits in general position bifurcate concurrently from this normal mode, we expect that the $y-$axis orbit remains stable after the bifurcation.
To verify this assert, we proceed as above by introducing the coordinates

\be
\left\{
  \begin{array}{cc}
    Y=& \sqrt{2J}\cos\theta \\
    P_Y= &\sqrt{2J}\sin\theta\\
    X=& X \\
    P_X=& U
  \end{array}
\right.
\ee
so that the periodic orbit is given by

\be
X=U=0,\;\;\;\;\;\; J=\frac52\E.\label{y-ay}
\ee
The system of differential equation for the perturbation of the normal mode is given by

\be
\frac{d}{dt}\left(
              \begin{array}{c}
                \d X \\
                \d U \\
              \end{array}
            \right)=\left(
                      \begin{array}{cc}
                        \widetilde {\mathcal K}_{UX} & \widetilde {\mathcal K}_{UU} \\
                       -\widetilde {\mathcal K}_{XX} & -\widetilde {\mathcal K}_{XU} \\
                      \end{array}
                    \right)\left(
              \begin{array}{c}
                \d X \\
                \d U \\
              \end{array}
            \right)\label{var_y}
\ee
where now $\widetilde {\mathcal K}=\mathcal K(X,U,\theta,J)$. Since  we are dealing with a perturbation of a 1:2 symmetric resonance, the terms proportional to $\cos(4\theta_1-2\theta_2)$ in $K_4$ are of second degree in $J_1$ and, as a consequence, the matrix of the second derivative of $\widetilde K$ computed on the $y-$normal mode does not depend on $\theta$. Thus,
we do not need to perform the transformation (\ref{compl_x}). The equations of variation (\ref{var_y}) give

\be
\frac{d}{dt}\left(
              \begin{array}{c}
                \d X \\
                \d U \\
              \end{array}
            \right)=
\left(
                      \begin{array}{cc}
                        \Omega_{11} & \Omega_{12} \\
                       \Omega_{21} & \Omega_{22} \\
                      \end{array}
                    \right)
 \left(
              \begin{array}{c}
                \d X \\
                \d U \\
              \end{array}
            \right)
\ee
where

\ba
\Omega_{11}&=&\Omega_{22}=0 \\
\Omega_{12}&=&-\Omega_{21}=1-\left(\frac{5 c}{8}\E+2 \tilde\delta\right)\e^2-\frac{25}{768} \E^2  \left(c (18 b+5 c)-48 c_1\right) \e^4.
\ea
Thus,

\ba
\det\Omega=\Omega_{12}^2&=&1+\left(-\frac{5 c \E}{4}+4 \tilde\delta \right) \e^2 +\frac{1}{384} \left(-450 b c \E^2+25 c^2 \E^2\right.\nn
&-&\left.960 c \E\tilde\delta +1536 \tilde\delta ^2+1200 \E^2 c_1\right) \e^4+O(\e^6)
\ea
has a positive zero order term, which implies that for $\e$ small enough the $y-$axis orbit is always stable.

\section{Application: 1:2 resonance in systems with elliptical equipotentials}

We now illustrate how to apply the above theory to some cases relevant for galactic dynamics. In particular, we are interested in  the question of the stability of axial orbits in triaxial potentials and of the possible existence of additional stable families of periodic orbits.
The present analysis of 2 DOF systems is a first step in this program because in allows us to study the dynamics in principal planes of triaxial systems with reflection symmetries. In turn these studies are useful to solve problems like the construction of self-consistent equilibria, the computation of isophotal shapes and velocity ellipsoids, etc.

We consider a fairly general class of potentials with self-similar elliptical equipotential and unit `core' radius of the form \cite{Bi,eva,MP}

\be
\label{pots}
{\mathcal V} (x,y;q,\alpha) =\left\{
                    \begin{array}{ll}
                      \frac{1}{\alpha}\left(1 + x^2 + \frac{y^2}{q^2}\right)^{\alpha/2},\;\; & 0<\alpha<2 \\
                      \frac12 \log\left(1 + x^2 + \frac{y^2}{q^2}\right),\;\; & \alpha = 0.
                    \end{array}
                  \right.
\ee
The physical parameters are $q$, the {\it ellipticity} of the equipotentials and $\alpha$, a shape parameter: on the two extremes of its range, $\alpha=0$ corresponds to the standard logarithmic potential and $\alpha=2$ to the anisotropic harmonic oscillator.

The family of potentials  (\ref{pots}) admits a  series expansion of the form
(\ref{seriesp})
with

\ba
V_0&=&  \frac12\left(x^2+ \frac{y^2}{q^2}\right), \\
V_2&=&  \frac{\alpha - 2}{8} \left(x^2+ \frac{y^2}{q^2}\right)^2,  \\
V_4&=&  \frac{(\alpha - 2)(\alpha - 4)}{48} \left(x^2+ \frac{y^2}{q^2}\right)^3 .
\ea
Since the unperturbed frequencies now are $\omega_1=1$ and $\omega_2=1/q$, we introduce the detuning parameter (\ref{DET12})

\be
\label{DEE12}
\delta \doteq q - \frac12.\ee
Performing the scalings  (\ref{scal_cc}) and (\ref{scal_det}) and
collecting  terms of the same order in $\e$  we obtain the Hamiltonian function (\ref{Has}), where the non-vanishing terms are now given by

\ba
H_0&=&\frac12(x_1^2+p_1^2)+(x_2^2+p_2^2) \label{ha0}\\
H_2&=&\tilde\d(x_1^2+p_1^2)- \frac{2-\alpha}{8} (x_1^2 + 2 x_2^2)^2  \label{ha2}\\
H_4&=&-\frac{\tilde\d}{4} (2-\alpha)(x_1^4 - 4 x_2^4) + \frac{(2-\alpha)(4-\alpha)}{48} \left(x_1^2 + 2 x_2^2\right)^3. \label{ha4}
\ea
Hence, in this particular case

\ba
a&=&\frac{1+ 2 \d}{2} (2-\alpha),\\
b&=&2(2-\alpha)(1- 2 \d),\\
c&=&2(2-\alpha),\\
a_1&=&\frac18(2-\alpha)(4-\alpha), \\
b_1&=&6a_1,\\
c_1&=&12a_1,\\
d_1&=&8a_1.
\ea
The presence of terms proportional to $\tilde\d$ in $H_4$ is due to the dependence of the potential \eqref{pots} on $q$.
However the existence and stability analysis of the periodic orbits of the system follows exactly the same way of the preceding section.

Since we have $12a+3b-4c=4(2-\alpha)>0$, the non degeneracy condition (\ref{prima_cond}) is satisfied and the system is able to exhibit the bifurcation of periodic orbits in general position. In order to establish their existence, we look at the sign of $6a-c$ and/or $3b-2c$. We find

\be
6a-c=(1+6 \d)(2-\alpha)  , \ee
thus for $\d>0$ these systems fall in the cases (\ref{bxp}) and (\ref{axp}). Hence banana
and anti-banana both bifurcate from the $x-$axis orbit respectively for $E=E_{B1}$ and $E=E_{A1}$, where
the thresholds in terms of the physical parameters are given in (\ref{E1B},\ref{E1A}) and for potential (\ref{pots}) turn out to be

\ba
E_{B1}&=& \frac{16}{2-\alpha }\left( q - \frac12 \right)+\frac{8 (41 \alpha-10 )}{3 (2-\alpha )^2}\left( q - \frac12 \right)^2 \label{E2BA} \\
E_{A1}&=&\frac{16 }{2-\alpha }\left( q - \frac12 \right)+\frac{8 (53 \alpha+14 )}{3 (2-\alpha )^2}\left( q - \frac12 \right)^2.\label{E2AA}
\ea
Since in this case we have

\be\label{mua}
\mu=3 \left(4-\alpha ^2\right)>0,\ee
in agreement with the general expression (\ref{GAB1}), the difference between the two thresholds is

\be
E_{A1}-E_{B1}=32\frac{2+\alpha}{(2-\alpha )^2}\left( q - \frac12 \right)^2.
\ee
 This relation and equations (\ref{detb}) and (\ref{deta}) establish the bifurcation of stable  banana orbits followed by unstable  anti-banana orbits. These results generalize those already obtained in the work on the logarithmic potential \cite{bbp,pbb} and provide good approximations to the numerical investigations available in literature \cite{mes,scu}.

\section{Conclusions}

We have presented the investigation of a fairly large class of natural reversible Hamiltonian systems close to the 1:2 resonance and endowed with reflection symmetry with respect to the configuration variables of both DOFs. By means of a resonant detuned normal form, we have obtained a general description of the bifurcation scenario of periodic orbits in general position (banana and anti-banana) from the normal modes.

 Equations (\ref{E1B}--\ref{E2B}) provide the energy levels for these bifurcation in terms of the detuning and the other physical parameters characterizing the system. We have found that the coefficients of the quartic term in the potential essentially determine the distinction between the bifurcations from either normal mode: our main result is that if banana and anti-banana orbits bifurcate from the $y-$normal mode, they do it at the same energy level; if instead the bifurcations are from the $x-$normal mode, their sequence (and the stability of the new orbits) is determined by the parameter $\mu$ defined in (\ref{mu}). We remark that for the reliability of the predictions based on these parameters it is essential that in the computation of the normal form terms of order four (degree six) in the potential are included. As an example, in the case of the `$\alpha-$models'  (\ref{pots}), if we arrest the expansion of the potential at order two and normalize up to order four, we get

\be\label{mub}
\mu=-9 \left(4-\alpha ^2\right)<0,\ee
obtaining a wrong prediction for the bifurcation sequence.

However, although these results are in excellent agreement with numerical simulations, we can not deduce from them the generic behavior of the system. For example, it is difficult to say if the concurrent bifurcation from the $y-$normal mode is persistent and, if not, at which order it split. Actually, the `catastrophe germ' for a 1:2 symmetric resonance is given by the second order term in the perturbation, but has infinite codimension \cite{Br1:1}. In simpler terms, this means that truncating at order four could not be enough and `a priori' one has to add (infinitely) many higher order terms to the series expansion (\ref{Has}) to obtain a faithful description of the true dynamics of the system. Additional efforts are therefore necessary for a full understanding of this problem.

\section*{Acknowledgments}

Our thanks to Giuseppe Gaeta and Heinz Han{\ss}mann for very fruitful discussions. This work is supported by INFN -- Sezione di Roma Tor Vergata and by the Scuola di Dottorato of the Dipartimento di Scienze di Base e Applicate per l'Ingegneria -- Universit\`a di Roma ``la Sapienza".

\end{document}